\documentclass[12pt]{article}
\usepackage{epsf}
\author{ A.\,Samsonov\thanks{e-mail:sams@heron.itep.ru}\\\\
\it {\small{Institute of Theoretical and Experimental Physics,}} \\
\it {\small{Bol'shaya Cheremushkinskaya, 25, Moscow, 117259 Russia}}}
\bf
\title{Magnetic moment of the $\rho$-meson in QCD sum rules}

\begin{document}
\date{}
\maketitle
\newcommand{\qq}{\langle\overline{q}q\rangle^2}

\def\la{\mathrel{\mathpalette\fun <}}
\def\ga{\mathrel{\mathpalette\fun >}}
\def\fun#1#2{\lower3.6pt\vbox{\baselineskip0pt\lineskip.9pt
\ialign{$\mathsurround=0pt#1\hfil##\hfil$\crcr#2\crcr\sim\crcr}}}

$$\centerline{\hbox{Abstract}}$$
\rm
\indent The magnetic moment $\mu$ of the $\rho$-meson 
is calculated in the framework of 
QCD sum rules in external fields.  
Bare loop calculations (parton model) give: $\mu_{part}=2.0$.
The contribution of operators of dimension 6 reduces this value: 
$\mu=1.5\pm 0.3$.
\\
\newpage
$\,$\\\\
\section{Introduction}
$\,$
\indent Investigation of the static properties of vector mesons provides
an important information about strong interaction of hadrons. In particular,
the vector dominance hypothesis (VDM) supposes that the interaction of real
or virtual photon with hadrons proceeds in such a way that the photon 
first transforms into vector mesons $\rho,\,\omega,\,\phi,$ which
then undergo interaction with hadrons. In the consistent lagrangian
formulation of VDM it is assumed \cite{kroll} (for review, see 
\cite{iof-kh-lip}) that $\rho$-mesons are 
Yang-Mills vector bosons. In the framework of this hypothesis
the $\rho$-meson magnetic moment is equal to 2 
(in units $eh/(2m_\rho c)$), at least if strong interaction is 
neglected. \\
\indent The goal of this paper is to calculate the $\rho$-meson magnetic 
moment in QCD, using the method of QCD sum rules in external fields
\cite{iof-sm-magmom},\cite{bal-yung}.\\
\indent In paper \cite{iof-sm-meson} the $\rho$-meson formfactors were found 
 at intermediate momentum transfer by QCD sum rules. By extrapolation
of the $\rho$-meson magnetic formfactor to the point $Q^2=0$ (outside
the applicability domain of the technique) it was found that the $\rho$-meson 
magnetic moment $\mu$ is close to 2. However, this result can not be 
considered as conclusive; the direct calculation of $\mu$ in QCD in model
independent way is still absent. The $\rho$-meson magnetic 
moment was calculated in the
Dyson-Schwinger equation based models \cite{hawes},\cite{hecht} and in
the framework of relativistic quantum mechanics \cite{rel-qm}.\\ 
\indent Here we work in the limit of zero quark masses, 
$\alpha_s$-corrections are neglected.

\section{Phenomenological part of the sum rule}
$\;$
\indent We consider the correlator of two vector currents in the external
electromagnetic field:
$$\Pi_{\mu\nu}(p)=i\int d^4x\,e^{ipx}\langle T(j_\mu(x)j^+_\nu(0)\rangle_F\,.
\eqno(1) $$
Here subscript $F$ denotes the presence of the external electromagnetic
field with strength $F_{\rho\lambda}$ and $j_\mu$ is the vector current 
with $\rho$-meson quantum numbers:
$j_\mu=\overline{u}\gamma_\mu d$. Its matrix element is 
$$\langle\rho^+\vert j_\mu \vert 0\rangle =(m_\rho^2/g_\rho)e_\mu\,,
\eqno(2)$$
where $m_\rho$ is the $\rho$-meson mass, $g_\rho$ is the $\rho$-$\gamma$ 
coupling constant, $g_\rho^2/(4\pi)=1.27$, and $e_\mu$ is the $\rho$-meson 
polarization vector. \\
\indent In the limit of weak external field we consider only linear in 
$F_{\rho\lambda}$ terms in the correlator $\Pi_{\mu\nu}$ (1):
$$\Pi_{\mu\nu}=\Pi^0_{\mu\nu}+i\sqrt{4\pi\alpha}
\Pi_{\mu\nu\chi\sigma}F_{\chi\sigma}\,.\eqno(3)$$
\indent We find magnetic moment from sum rule for the invariant  function 
$\Pi(p^2)$ at certain kinematical structure of $\Pi_{\mu\nu\chi\sigma}$ (3). 
To obtain this sum rule, we calculate $\Pi$ at $p^2<0$ as the 
operator product expansion series. On the other hand, we saturate 
dispersion relation for $\Pi$ by the contributions of physical
states.
After equating of these representations the required sum rule
appears. \\ 
\indent Therefore, first of all one should choose kinematical structure. \\
\indent The electromagnetic vertex of the $\rho$-meson has the following
general form \cite{iof-sm-meson}:
$$\displaylines{\langle\rho(p+q,e^{r^\prime})|j^{el}_\chi|\rho(p,e^r)\rangle =
-e^{r^\prime}_\sigma e^r_\rho \bigg(\Big((2p+q)_\chi g_{\rho\sigma}-
(p+q)_\rho g_{\chi\sigma}-p_\sigma g_{\rho\chi}\Big)F_1(-q^2)+\hfill}$$
$$\displaylines{\hfill+(g_{\chi\rho}q_\sigma-g_{\chi\sigma}q_\rho)
F_2(-q^2)+{1\over{m_\rho^2}}
(p+q)_\rho p_\sigma(2p+q)_\chi F_3(-q^2)\bigg)\,.~~~~(4)}$$
In (4) $j^{el}_\chi=e_u\overline u\gamma_\chi u+e_d\overline d\gamma_\chi d$
is electromagnetic current, $e_u\,,e_d$ are $u$- and $d$-quark charges and
$F_1\,,F_2\,,F_3$ are electric, magnetic and quadrupole formfactors
correspondingly, 
$$F_1(0)=1\,,~~~~\mu=1+F_2(0)\,,\eqno(5)$$
$\mu$ is the $\rho$-meson magnetic moment. \\
\indent Using (2) and (4), we obtain for the 
$\langle 0|j_\mu|\rho\rangle\langle
\rho|j_\chi|\rho\rangle\langle\rho|j_\nu|0\rangle\epsilon_\chi$
transition:
$$\displaylines{
-i\sum\limits_{r,r^\prime}\langle 0|j_\mu|\rho^{r^\prime}\rangle
\langle\rho^{r^\prime}|j^{el}_\chi|\rho^r\rangle\langle\rho^r|
j_\nu|0\rangle\epsilon_\chi=\hfill}$$
$$\displaylines{\hfill=i{m_\rho^4\over{g_\rho^2}}\sum_{r,r^\prime}
 e^{r^\prime}_\mu e^{r^\prime}_\sigma e^r_\rho e^r_\nu \epsilon_\chi
\bigg(\Big((2p+q)_\chi g_{\rho\sigma}-
(p+q)_\rho g_{\chi\sigma}-p_\sigma g_{\rho\chi}\Big)F_1(-q^2)+
~~~~~~(6)}$$
$$\displaylines{\hfill+(g_{\chi\rho}q_\sigma-g_{\chi\sigma}q_\rho)
F_2(-q^2)+{1\over{m_\rho^2}}
(p+q)_\rho p_\sigma(2p+q)_\chi F_3(-q^2)\bigg)\,.~~~~}$$
Here  $\epsilon_\chi-$photon polarization, $r,\,r^\prime$ are 
the $\rho$-meson polarization indices. Let us consider in this 
expression linear in $q_\sigma$ terms. We sum over 
$\rho$-meson polarizations, retain the antisymmetric over $\chi,\sigma$ part,
introduce $F_{\chi\sigma}=
i(\epsilon_\chi q_\sigma-\epsilon_\sigma q_\chi)$ and obtain for (6):
$$\displaylines{-{m_\rho^4\over{2g_\rho^2}}F_{\chi\sigma}
\bigg((F_2+{1\over 2}F_1){1\over{p^2}}\Big(p_\nu
(p_\chi g_{\mu\sigma}-p_\sigma g_{\mu\chi})-p_\mu
(p_\chi g_{\nu\sigma}-p_\sigma g_{\nu\chi})\Big)+\hfill}$$
$$\displaylines{\hfill+{1\over 2}F_1{1\over{p^2}}\Big(p_\nu
(p_\chi g_{\mu\sigma}-p_\sigma g_{\mu\chi})+p_\mu
(p_\chi g_{\nu\sigma}-p_\sigma g_{\nu\chi})\Big)+(F_2+F_1)
(g_{\mu\chi}g_{\nu\sigma}-g_{\mu\sigma}g_{\nu\chi})\bigg)\,.}$$
Formfactor $F_3$ does not give linear in $q_\sigma$ contribution. \\
\indent Thus, we choose the structure
$$p_\nu(p_\chi g_{\mu\sigma}-p_\sigma g_{\mu\chi})-
p_\mu(p_\chi g_{\nu\sigma}-p_\sigma g_{\nu\chi})\,.\eqno(7)$$
In comparison with another
possible structure, $g_{\mu\chi}g_{\nu\sigma}-
g_{\mu\sigma}g_{\nu\chi}$, (7) contains two additional powers of momentum in
the numerator, which result in better convergence of the operator expansion
series. \\
\indent It should be noted here that,
as follows from the vector current 
conservation, the antisymmetric over field indices $\chi,
\sigma$ structure in $\Pi_{\mu\nu\chi\sigma}$ (3) is 
antisymmetric over $\rho$-meson indices $\mu,\nu$ too. \\  
\indent From (5) one can see that
$F_2(0)+1/2F_1(0)=\mu-1/2$. \\
\indent Thus, one should calculate the invariant function $\Pi(p^2)$ 
at the structure (7) in $\Pi_{\mu\nu\chi\sigma}$. In the dispersion relation
for $\Pi$ we use the simplest model of physical spectrum, which contains 
the lowest resonance and continuum.
Phenomenological representation of $\Pi$ has
the form:
$$\Pi(p^2)=\int ds{\rho_L(s)\over{(s-p^2)^2}}+...
\,,$$
$$\rho_L(s)=-{m_\rho^4\over{2g_\rho^2}}{1\over s}\Big(\mu-{1\over 2}\Big)
\delta(s-m_\rho^2)+
f(s)\theta(s-s_\rho)\,.$$
Here dots mean the contributions of non-diagonal 
transitions (for example, \\ 
$\langle 0|j_\mu|\rho^\star\rangle\langle
\rho^\star|j_\chi \epsilon_\chi|\rho\rangle\langle\rho
|j_\nu|0\rangle$, where $\rho^\star$ is the excited state with 
the same quantum numbers as $\rho$), 
function $f$ represent continuum contribution 
and $s_\rho$ is the continuum threshold for the $\rho$-meson.\\
\indent Retaining only the terms, 
which do not vanish after Borel transformation, we obtain:
$$\Pi(p^2)=-{m_\rho^2\over{2g_\rho^2}}
{{\mu-{1\over 2}}\over{(m_\rho^2-p^2)^2}}+
{\tilde{C}\over{m_\rho^2-p^2}}+
\int\limits_{s_\rho}^\infty ds{f(s)\over{(s-p^2)^2}}\,,\eqno(8)$$
where $\tilde{C}$ appears due to non-diagonal transitions.
\section{Calculation of the vector current correlator}
$\,$ \indent Now let us calculate $\Pi(p^2)$, basing on the
operator product expansion in QCD. \\
\indent The quark propagator
in the external electromagnetic field $F_{\mu\nu}$ in the fixed-point
gauge $x_\mu A_\mu=0\,,\;A_\mu=-1/2F_{\mu\nu}x_\nu$ can be found in
\cite{iof-sm-magmom}:
$$\displaylines{\langle Tq_\alpha^a(x)\overline{q}_\beta^b(0)\rangle_F=
{i\delta^{ab}(\hat{x})_{\alpha\beta}\over{2\pi^2x^4}}-
{\delta^{ab}g_{\alpha\beta}\over{12}}\langle\overline{q}q\rangle-
{{i\delta^{ab}\langle\overline{q}\sigma_{\rho\lambda}q\rangle_F}\over{48}}
(\gamma_\rho \gamma_\lambda-\gamma_\lambda\gamma_\rho)_{\alpha\beta}-
\hfill}$$
$$\displaylines{\hfill-
{{\delta^{ab}\sqrt{4\pi\alpha}e_q F_{\rho\lambda}}\over{32\pi^2 x^2}}
(\hat{x}\gamma_\rho \gamma_\lambda+
\gamma_\rho \gamma_\lambda\hat{x})_{\alpha\beta}\,.}$$
Here $e_q$ is the quark charge, $\alpha,\,\beta$ are spinor indices, 
$a,b$ - color indices and (see \cite{iof-sm-magmom})
$\langle\overline{q}\sigma_{\rho\lambda}q\rangle_F=
\sqrt{4\pi\alpha}e_q \chi F_{\rho\lambda}\langle\overline{q}q\rangle\,,$
$\chi$ is the quark condensate magnetic susceptibility. \\
\indent The expression for the quark propagator
in the external electromagnetic and soft gluon fields has the
following form in momentum representation \cite{iof-sm-magmom}:
$$\displaylines{\hat{S}_{FG}=
-{{ige_q \sqrt{4\pi\alpha}F_{\rho\lambda}G^n_{\sigma\tau}t^n}
\over{2p^6}}\Big(\gamma_\lambda \gamma_\tau \gamma_\rho \gamma_\sigma \hat{p}
-2p_\lambda \gamma_\tau \gamma_\sigma \gamma_\rho-
 2p_\tau \gamma_\lambda \gamma_\rho \gamma_\sigma-\hfill}$$
$$\displaylines{\hfill{{-8p_\rho p_\tau g_{\lambda\sigma}}\over{p^2}}\hat{p}
+2g_{\lambda\tau}\gamma_\sigma  \gamma_\rho \hat{p}
-2g_{\lambda\tau}g_{\rho\sigma}\hat{p}\Big)\,.}$$
Here $G^n_{\sigma\tau}$ is the gluon field strength and $t^n$ are the color
matrices. \\
\indent  The contribution of
the loop diagrams to $\Pi(p^2)$ is equal to
$$-{3\over{16\pi^2}}\int\limits_0^\infty{ds\over{(s-p^2)^2}}\,.\eqno(9)$$
According to the quark-hadron duality, the continuum contribution in the
interval of $P^2=-p^2$ from $s_\rho$ to infinity is determined by the
bare loop in this interval. 
Therefore, function $f$ in (8) is constant: $f=-3/(16\pi^2)$.\\
\indent The loop diagrams correspond to the operator of the lowest dimension
$F_{\rho\lambda}$. Operators of dimension 4 are absent. As was shown in 
\cite{iof-sm-magmom}, operator $\overline{q}(D_\mu \gamma_\nu -
D_\nu \gamma_\mu)q$ has opposite with respect to the electromagnetic field
C-parity and can not be induced by them, while operator
$\epsilon_{\mu\nu\rho\lambda}\overline{q}\gamma_5 \gamma_\rho D_\lambda q$
vanishes due to equation of motion for massless quarks. \\
\indent There are a number of vacuum expectation values of operators of
dimension 6:\\
$\langle\overline{q}\sigma_{\rho\lambda}q\rangle_F
\langle\overline{q}q\rangle$,
$\langle G^n_{\sigma\tau}G^n_{\sigma\tau}\rangle F_{\mu\nu}$
and
$$g\langle\overline{q}\Big((G^n_{\mu\lambda}D_\nu-
\overleftarrow{D}_\nu G^n_{\mu\lambda})-(G^n_{\nu\lambda}D_\mu-
\overleftarrow{D}_\mu G^n_{\nu\lambda})\Big)
\gamma_\lambda t^n q\rangle_F\,,$$
$$\epsilon_{\mu\nu\rho\lambda}g\langle\overline{q}(G^n_{\rho\xi}D_\lambda+
\overleftarrow{D}_\lambda G^n_{\rho\xi})
\gamma_\xi \gamma_5 t^n q\rangle_F\,,\eqno(10)$$
$$d^{ikl}(G^i_{\mu\lambda}G^k_{\lambda\rho}G^l_{\rho\nu}-
G^i_{\nu\lambda}G^k_{\lambda\rho}G^l_{\rho\mu})\rangle_F\,,$$
where $D_\mu$ is the covariant derivative and $d^{ikl}$ are SU(3) structure 
constants. \\
\indent The diagrams, corresponding to the operator
$\langle G^n_{\sigma\tau}G^n_{\sigma\tau}\rangle F_{\mu\nu}$, 
have infrared divergence. We introduce the cut-off over 
transversal momenta $\lambda$ and obtain their contribution into
$\Pi(p^2)$:
$$-{1\over{36}}\langle{\alpha_s\over\pi}G^2\rangle
\Big({1\over{2\lambda^4p^2}}-{1\over{6\lambda^2p^4}}+{3\over p^6}\Big)\,.
\eqno(11)$$
This divergence is probably cancelled by the contribution of field induced
vacuum expectation values (10) (for example, this was shown in \cite{og-sams}
for dimension 4 operators and symmetric tensor field). Usually such vacuum
expectation values can be calculated by constructing corresponding sum rule.
But for operators (10) this approach is inapplicable because of their high
dimension. \\
\indent The dominating contribution appears from no loop diagrams with
hard gluon exchange. In our case such diagrams contain the operator
$\langle\overline{q}\sigma_{\rho\lambda}q\rangle_F
\langle\overline{q}q\rangle$. They give:
$${2\over 9}{g^2\qq\chi\over{p^6}}\,.\eqno(12)$$
It should be noted here that the quark condensate magnetic 
susceptibility $\chi$ is negative. \\
\indent Collecting the expressions (9), (11), (12), one can find the operator
product expansion part of the sum rule:
$$\Pi(p^2)=-{3\over{16\pi^2}}\int\limits_0^\infty{ds\over{(s-p^2)^2}}
+{2\over 9}{g^2\qq\chi\over{p^6}}
-{1\over{36}}\langle{\alpha_s\over\pi}G^2\rangle
\Big({1\over{2\lambda^4p^2}}-{1\over{6\lambda^2p^4}}+{3\over p^6}\Big)\,.
\eqno(13)$$
The uncertainty in the contribution of the vacuum expectation values (10) is 
included into the errors.
\section{Results and discussion}
\indent    
After Borel transformation 
$$\hat{B}(M^2)=\lim_{P^2,n \to \infty \atop{P^2/n=M^2}}
{(P^2)^{n+1}\over{n!}}\bigg(-{d\over{dP^2}}\bigg)^n\,,~~~~P^2=-p^2>0$$
we equate the phenomenological (8) and operator product expansion (13) parts
of sum rule and obtain:
$$\displaylines{\mu-{1\over 2}+CM^2={3g_\rho^2M^2\over{8\pi^2m_\rho^2}}
\Big(1-e^{-s_\rho/M^2}\Big)e^{m_\rho^2/M^2}-\hfill}$$
$$\displaylines{\hfill-{g_\rho^2\over{m_\rho^2}}e^{m_\rho^2/M^2}
\Bigg(-{2g^2\qq\chi\over{9M^2}}+
{1\over 36}\langle{\alpha_s\over\pi}G^2\rangle\Big({M^2\over{\lambda^4}}+
{1\over{3\lambda^2}}+{3\over{M^2}}\Big)\Bigg)\,.~~~~(14)}$$
$C$ appears due to nondiagonal transitions. \\
\indent We use the following values of parameters: \\
$m_\rho=0.77\,GeV-$the $\rho$-meson mass, \\
$g_\rho^2/(4\pi)=1.27-$the $\rho$-$\gamma$ coupling constant, \\
$s_\rho=1.5\,GeV^2-$the continuum threshold for $\rho$-meson,\\
$\langle(\alpha_s/\pi)G^2\rangle=0.009\pm 0.007\,GeV^4-$the gluon condensate  
\cite{iof0207191}, \\
$g^2\qq=(0.28\pm 0.09)\times 10^{-2}\,GeV^6-$the quark condensate 
\cite{iof0207191}, \\
$\chi=-(5.7\pm 0.6)\,GeV^{-2}-$the quark condensate magnetic 
susceptibility \cite{bel-kog},\\
$\lambda^2=0.8\,GeV^2-$the cut-off over transversal momenta. \\
\indent First of all, let us consider the contribution of the bare loop (and
continuum). It is given by the first term in (14). In \cite{svz} the 
following relation for $g_\rho$ can be found:
$${g_\rho^2M^2\over{4\pi^2m_\rho^2}}\Big(1-e^{-s_\rho/M^2}\Big)
e^{m_\rho^2/M^2}=1\,.\eqno(15)$$
Substituting (15) into (14) and omitting for a while the terms with quark
and gluon condensates, one can obtain very simple answer:
$$\mu-{1\over 2}={3\over 2}\,.$$
We see that in the parton model approximation the $\rho$-meson magnetic 
moment is equal to 2. This result agrees with prediction of the 
vector dominance hypothesis. \\
\indent Now let us analyze the whole equation (14). 
In order to find the value of magnetic moment, we approximate the right-hand 
side of (14) (see fig.1) by a straight line in the  interval
$0.9\,GeV^2 \le M^2\le 1.3\,GeV^2$ and find it ordinate
at zero Borel mass. \\ 
\indent Thus we obtain:
$$\mu=1.5\,.\eqno(16)$$
\indent The contribution of the operators of dimension 6 to this value 
does not exceed 20\%. \\
\indent
The contribution of the terms, which contain $\lambda^2$, 
is not more than $20\%$
of the total contribution of dimension 6 operators. That is why variation of 
$\lambda^2$ within the interval $0.6\,GeV^2 \le \lambda^2\le 1.0\,GeV^2$ 
does not change the value of magnetic moment. \\
\indent The variations of the values of the quark and gluon condensates
within the given limits change the value of magnetic moment by 
$\la 10\%$ each. \\ 
\indent The uncertainty in the value of the quark condensate magnetic 
susceptibility results in the error about few percent 
in the value of the magnetic
moment. Variation of the continuum threshold for the $\rho$-meson $s_\rho$ 
in the reasonable limits gives the same effect. \\ 
\indent Supposing that the contribution of vacuum
expectation values (10) does not exceed $50\%$ of that from diagrams with
hard gluon exchange (12), we obtain after collecting all uncertainties: 
$$\mu=1.5\pm 0.3\,.$$
This is our final result. \\
\indent In \cite{iof-3vertex} it was shown that the approximation procedure
is correct (nonlinear terms can be safely neglected), when $\mu\gg CM^2$.
In our case $CM^2/\mu\approx 0.2\div 0.3$.\\
\indent Thus we find that in parton model (bare loop) approximation 
the  $\rho$-meson magnetic moment $\mu_{part}=2$, 
whereas the nonperturbative interactions decrease this quantity by a quarter:
$\mu=1.5\pm 0.3$. It is important to mention that all accounted operator
product expansion corrections are negative, i.e. results in decrease of $\mu$
in comparison with $\mu_{part}=2$. Since the effective values of the
Borel parameter $M^2$ are about $1\,GeV^2$, one may expect that 
perturbative corrections are remarkable and can reach $\sim 20\%$.\\
\indent The value of the $\rho$-meson magnetic moment was calculated in a
number of papers within the Dyson-Schwinger equation 
based models. In 
\cite{hawes} the value $\mu=2.69$ was found. In \cite{hecht} 
several results are compared, and the values of $\mu$ 
lie between 2.5 and 3.0. Relativistic quantum mechanics model
gives \cite{rel-qm} $\mu=2.23\pm 0.13$.
Unfortunately, while $\alpha_s$-corrections are not calculated (we plan to 
do it in the next paper), it is hard to say with certainty if this
discrepancy is real or not. 
\\ \\
\indent The author is grateful to B.L.\,Ioffe for posing the problem and 
valuable discussions and to A.G.\,Oganesian for helpful discussions. \\
\indent The work is supported in part by grants CRDF RP2-2247, INTAS 2000
Project 587 and RFFI 00-02-17808.

\begin{figure}[h]
\epsfxsize=12.0cm
\epsfbox{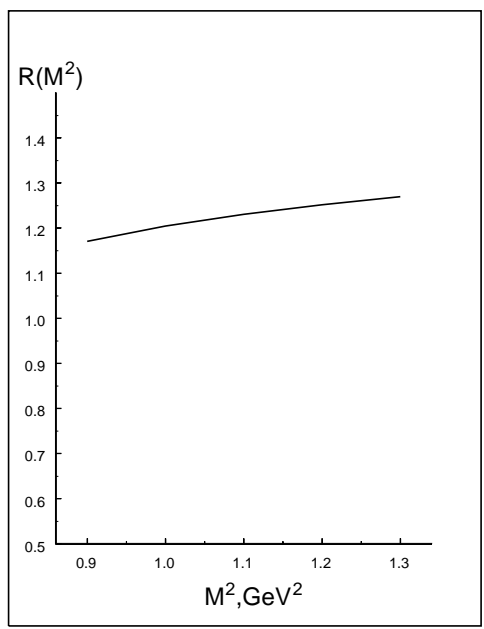}
\caption{The right-hand side of equation (14) $R(M^2)$ 
as the function of $M^2$.}
\end{figure}
\end{document}